\documentclass[aps,preprint]{revtex4}
\usepackage[dvips]{graphicx}
\usepackage{amssymb}
\usepackage{color}

\begin{document}
\draft

\title{Application of topological resonances in experimental investigation of a Fermi golden rule in microwave networks}

\author{Micha{\l} {\L}awniczak,$^{1}$ Ji\v{r}\'{\i} Lipovsk\'{y},$^{2}$ Ma{\l}gorzata Bia{\l}ous,$^{1}$ and Leszek Sirko$^{1}$}
\address{$^{1}$Institute of Physics, Polish Academy of Sciences, Aleja  Lotnik\'{o}w 32/46, 02-668 Warszawa, Poland\\
$^{2}$Department of Physics, Faculty of Science, University of Hradec Kr\'alov\'e, Rokitansk\'eho 62, 500 03 Hradec Kr\'alov\'e, Czechia\\
}
\date{\today}

\begin{abstract}
We investigate experimentally a Fermi golden rule in  two-edge and five-edge microwave networks with preserved time reversal invariance. A Fermi golden rule gives rates of decay of states obtained by perturbing embedded eigenvalues of graphs and networks. We show that the embedded eigenvalues are connected with the topological resonances of the analyzed systems and we find the trajectories of the topological resonances in the complex plane.
\end{abstract}


\bigskip
\maketitle

\section{Introduction}
There are many processes in physics that can be described as a perturbation of a certain, usually quite symmetric system. One example of this behaviour are the eigenvalues of quantum systems, which after the perturbation of the initial Hamiltonian become resolvent resonances. A simple system in which this phenomenon can be studied is the model of quantum graphs (see \cite{BerkolaikoKuchment13}) -- a net-like structure equipped with the Hamiltonian of a single quantum particle. One can consider an infinite quantum graph, consisting of internal (finite) edges and external (infinite) edges. For rationally related lengths of the internal edges, certain eigenfunctions corresponding to eigenvalues embedded into the continuous spectrum can be constructed as sine functions on some of the edges with zeros at the vertices and vanishing on the infinite edges. However, if the rationality of the edge lengths is perturbed, the former eigenvalues travel into the complex plane and become resonances. Due to the topological nature of the former eigenstates, these resonances are usually in the literature called \emph{topological resonances}. The behaviour of the topological resonances and their trajectories near the initial eigenvalue have been of interest to many papers recently \cite{ExnerLipovsky10,GnutzmannSchanzSmilansky13,LeeZworski16,ColindeVerdiereTruc18}. In \cite{ExnerLipovsky10}, the trajectories of topological resonances depending on the edge lengths were found and the circumstances under which the resonance again becomes an eigenvalue were studied. In \cite{GnutzmannSchanzSmilansky13} the term ``topological resonances'' was first used and the statistical properties of these resonances were investigated. The paper \cite{ColindeVerdiereTruc18} proved that for tree graphs with at most one vertex of valency 1, the resonances are far from the real axis (and hence there are no ``narrow'' resonances), which is not the case for the other types of graphs.

In the paper \cite{LeeZworski16}, the authors pointed out the link between the so-called Fermi golden rule in physics and the speed with which the former eigenvalue moves in the complex plane. To be more precise, if the lengths of the edges are parametrized by a parameter $t$, they found a formula for $\mathrm{Im}\,\ddot k$, the imaginary part of the second derivative of the wave vector~$k$ (the square root of the complex energy of the resonance) with respect to the parameter~$t$. Another version of the formula, obtained by Lee and Zworski in the case of standard (Kirchhoff's) coupling conditions, was later found in \cite{ExnerLipovsky17} using the pseudo-orbit expansion method for general coupling conditions.

In the present paper, we put under the experimental test the results of Lee and Zworski \cite{LeeZworski16} using microwave networks. We find the trajectories of the resonances for two particular networks and compare them with the numerical simulations. As opposed to  abstract open quantum graphs microwave networks \cite{Hul2004} are real-world  open systems which are additionally characterized by intrinsic absorption. Yet, we find a good correspondence of the experimental and theoretical results; the experimental trajectories match the theoretical ones \cite{LeeZworski16}. Moreover, we verify a Fermi golden rule by computing $\mathrm{Im}\,\ddot k$ theoretically and comparing it with the fit of the experimental data. We find a good correspondence for both systems.

\section{Theoretical model and formulation of a Fermi golden rule}\label{sec2}
In this section, we introduce the well-known model of quantum graphs. Since the telegraph equation for microwave networks has a similar form as the Schr\"odinger equation for quantum graphs, our experiment can very precisely simulate the theoretical predictions made for quantum graphs.

Let us consider a metric graph consisting of a set of vertices $\mathcal{V}$ and a set of edges $\mathcal{E}$ including $M$ external (infinite) edges, which are parametrized by the intervals $(0,\infty)$ and denoted by $\{e_s\}_{s = 1}^M$, $s = 1,\dots, M$ and $N$ internal (finite) edges, which can be parametrized by the intervals $(0,\ell_j)$, $j = M+1,\dots, M+N$, connect two vertices and are denoted by $\{e_j\}_{j = M+1}^{M+N}$. We can denote $\ell_s  = \infty$ for $s = 1, \dots, M$. In the Hilbert space $\mathcal{H} = \oplus_{j = 1}^{M+N} L^2(0,\ell_j)$ we define the operator $H$ acting as the negative second derivative on the domain consisting of functions belonging to the Sobolev space $\oplus_{j = 1}^{M+N} W^{2,2}(0,\ell_j)$ and satisfying the standard (Kirchhoff's) coupling conditions at each vertex $v$, namely the continuity of the function values and vanishing of the sum of the derivatives
\begin{equation}
  u_i(v) = u_j(v)\,,\ v \in e_i\cap e_j\,,\quad \sum_{e_j \ni v} \partial_\nu u_j(v) = 0\,,\label{eq:cc}
\end{equation}
where $\partial_\nu u_j(v)$ is the derivative at the vertex $v$ in the direction into the vertex: $\partial_\nu u_j(0) = -u_j'(0)$, $\partial_\nu u_j(\ell_j) = u_j'(\ell_j)$. For more details on quantum graphs we refer the reader to the monograph \cite{BerkolaikoKuchment13}.

Before stating the main formula, let us define one of its components, the generalized eigenfunctions. Let $e^s(k,x)$, where $k$ is the square root of energy, $k^2 = E$, $s=1,\dots, M$ be for $k^2$ not belonging to the spectrum of $H$ characterized as follows
\begin{enumerate}
 \item $e^s(k,x)$ locally belongs to the domain of $H$,
 \item $(H-k^2) e^s(k,x) = 0$,
 \item $e^s_j(k,x) = \delta_{js} \mathrm{e}^{-\mathrm{i}kx}+ s_{js}\mathrm{e}^{\mathrm{i}kx}$, $j = 1,\dots, M$, where $e^s_j$ is the component of $e^s$ on the $j$-th infinite edge and $\delta_{js}$ is the Kronecker delta,
 \item We holomorphically extend $e^s_j(k,x)$ to all $k\in \mathbb{R}$.
\end{enumerate}

A Fermi golden rule  can be formulated as follows \cite[Thm. 1]{LeeZworski16}:

Let $H(t)$ be the above defined Hamiltonian, for which the lengths of the internal edges depend on the parameter $t$ via $\ell_j (t) = \ell_j\mathrm{e}^{-a_j(t)}$, $a_j(0) = 0$. Let $\dot a_j = \left.\frac{\partial a_j(t)}{\partial t}\right|_{t=0}$, $j = M+1,\dots M+N$ be the components of $\dot a = \{\dot a_j\}_{j=1}^M$. Let $k^2>0$ be a simple eigenvalue of $H(t)$ for $t=0$ and let $u$ be its corresponding normalized eigenfunction. Then there is a smooth function $k(t)$, where $k(0) = k$ is the eigenvalue and $k^2(t)$, $t\ne 0$ are resolvent resonances. For the second derivative $\ddot k$ of $k(t)$ with respect to $t$ at $t=0$, it holds
\begin{equation}
 \mathrm{Im}\, \ddot k = -\sum_{s=1}^M  |F_s|^2\,,\\
 \end{equation}
 where
 \begin{equation}
 F_s = k \left<\dot a u(x),e^s(k, x)\right>+ \frac{1}{k}\sum_v\sum_{e_j\ni v}\frac{1}{4}\dot a_j[3\partial_\nu u_j(v) \overline{e^s(k, v)}-u(v)\partial_\nu\overline{e^s_j(k,v)}]\,.
\label{Eq:F_s}
\end{equation}
In Eq. [\ref{Eq:F_s}] the inner product on $\mathcal{H}$  is denoted by $\left<\cdot, \cdot\right>$  and the overline denotes the complex conjugation.

In order to demonstrate the connection of this formulation of a Fermi golden rule to its usual form in the physics literature
let us consider a perturbation $H(t) = H_0+tV$ of the Hamiltonian $H_0$. Let $E_0 = E(0)$ be a simple eigenvalue of $H_0$ and let $E(t) = \sum_{n=0}^\infty a_n t^n$ be the complex resonance for the operator $H(t)$. Then the second derivative of $ \Gamma(t)  = 2 \,\mathrm{Im\,}E(t)$
can be expressed by an integral expression (18.3) in Ref. \cite{Lipovsky16}. Moreover, $\Gamma(t)$ can be related to the transition probability between the initial and final state, which is connected to the usual definition of a Fermi golden rule (for details see \cite{ReedSimon78}).

\section{Simulation of quantum graphs by microwave networks}\label{sec3}

Quantum graphs can be considered as idealizations of physical networks in the
limit where the widths of the wires are much smaller than their lengths. They can be
 successfully used to  model theoretically a broad range of physical problems, see, e.g.,
\cite{Gnutzmann2006}. From the experimental point of view the recent epitaxial techniques allow for designing and fabrication of  relatively simple quantum nanowire graphs  \cite{Samuelson2004,Heo2008}. However, to deal experimentally with much more complex systems characterized by many controllable parameters one has to use microwave networks.   Hul et al.\ \cite{Hul2004} demonstrated that quantum graphs with preserved time ($T$) reversal invariance
can be successfully simulated by microwave networks containing microwave junctions and coaxial cables.  In the investigations presented here the microwave networks were built of  microwave T-junctions and the  SMA-RG402 coaxial cables. The SMA-RG402 cables consist of two conductors. The inner conductor of a cable  with radius $r_1=0.05$~cm is surrounded by a concentric conductor of inner radius $r_2=0.15$~cm. The space between the conductors is filled with Teflon with a dielectric constant $\varepsilon\simeq 2.06$. For the specified  parameters of the cables  inside  them only the fundamental TEM mode propagates below the cut-off frequency of the TE$_{11}$ mode $\nu_{c}\simeq\frac{c}{\pi (r_1+r_2)\sqrt{\varepsilon}} \simeq 33$~GHz~\cite{Savytskyy2001}, where $c$ is the speed of light in vacuum.  The lengths of the edges in the corresponding quantum graph are defined by the optical lengths $\ell_i=\ell^g_i\sqrt{\varepsilon}$, where  $\ell^g_i$ are the geometric lengths of the coaxial cables.

 Microwave networks provide a very rich platform for the experimental and the theoretical study of quantum graphs possessing the same topology and boundary conditions at the vertices. The spectral and scattering properties of microwave networks have
been studied in Refs. \cite{Hul2004,Hul2005,Hul2005a,Lawniczak2008,Lawniczak2010,Lawniczak2012,Sirko2016,Bialous2016,Stockmann2016,Dietz2017}.  The microwave networks allow for simulations of variety of chaotic systems whose spectral properties can be described by the three main symmetry classes:  Gaussian orthogonal ensemble (GOE) \cite{Hul2004,Sirko2016,Lawniczak2019}, Gaussian unitary ensemble (GUE) \cite{Hul2004,Lawniczak2010,Bialous2016,Lu2020,Yunko2020} and Gaussian symplectic ensemble (GSE) \cite{Stockmann2016,Lu2020} in the Random Matrix Theory.

In this way microwave networks have become another important model systems to which belong  flat microwave cavities  \cite{Stockmann1990,Sridhar1994,Sirko1997,Hlushchuk2000,Hlushchuk2001,Blumel2001,Dhar2003,HemmadyPRL2005,Hul2005,Hemmady2006,Dietz2015,BialousPRE2019,Dietz2019} and  experiments using Rydberg atoms strongly driven by microwave fields \cite{Blumel1991,Jensen1991,Bellerman1992,Sirko1993,Buchleitner1993,SirkoPRL1993,Bayfield1995,Sirko1995,Sirko1996,Bayfield1999,Sirko2001,Sirko2002,Galagher2016} that are successfully used in simplifying experimental analysis of complex quantum systems.

In order to test experimentally a Fermi golden rule in microwave networks we consider two examples of quantum graphs shown in Fig.~1(a) and Fig.~1(c). A two-edge graph in Fig.~1(a) consists two vertices, two internal edges and two infinite leads. The second graph, a five-edge graph in Fig.~1(c), is more complex. It contains 4 vertices, five internal edges and two infinite leads.
The corresponding microwave networks constructed from microwave
coaxial cables are shown in Fig.~1(b) and Fig.~1(d).

\section{Theoretical results for a Fermi rule}\label{sectheor}
\subsection{A two-edge graph}

Let us consider a graph consisting of two vertices, two internal and two external edges (see Fig.~1(a)). Let the lengths of the internal edges $e_3$ and $e_4$ be $\ell_3<\infty$ and $\ell_4 <\infty$, respectively, while the edges $e_1$ and $e_2$ have infinite lengths. We will consider the dependence of the edge lengths on the parameter $t$ as $\ell_3 = \ell_0(1-t)$, $\ell_4 = \ell_0$ and the eigenvalue for $t=0$ with $k=\frac{2\pi}{\ell_0}$.
In the appendix we prove that for a two-edge graph a Fermi golden rule is expressed by the formula

\begin{equation}
  \mathrm{Im\,}\ddot k = -\frac{\pi^2}{2\ell_0}\,.
\end{equation}
Furthermore, we show  that the imaginary part of $k(t)$ near the eigenvalue behaves as
\begin{equation}
  \mathrm{Im}\,k \approx -\frac{\pi^2}{4\ell_0} t^2\,.
\end{equation}

\subsection{A five-edge graph}

Let us consider a graph in Fig.~1(c), having five internal edges and two external edges. Let the edge lengths be $\ell_3 = \ell_0(1-t)$, $\ell_4 = \ell_0(1+t)$, $\ell_5 = \ell_0(1-t)$, $\ell_6 = \ell_0(1+t)$, $\ell_7 = \ell_0(1+t)$ (this corresponds to the case \cite[Figure 4 c)]{LeeZworski16}).  Let us start from the eigenvalue with $k\ell_0 = \arccos{(-1/3)} = 1.9106$. For our choice we have
\begin{equation}
  \dot a_3 = 1\,,\quad \dot a_4 = -1\,,\quad \dot a_5 = 1\,,\quad \dot a_6 = -1\,.
\end{equation}
The computation of $\mathrm{Im\,}\ddot k$ is given in \cite[Sec. 18.2]{Lipovsky16} and a Fermi golden rule takes the form
\begin{equation}
  \mathrm{Im\,}\ddot k = -\frac{1}{\ell_0}[(\dot a_3-\dot a_6)^2+(\dot a_4-\dot a_5)^2]0.1711-\frac{1}{\ell_0}(\dot a_3-\dot a_6)(\dot a_4-\dot a_5)0.1141\,=-\frac{0.9124}{\ell_0}.
\end{equation}

\section{Experimental results}

Both microwave networks shown in Fig.~1(b) and Fig.~1(d)  can be described in terms of $2\times 2$ scattering matrix
$\hat S(\nu)$:
\begin{equation}
\label{eq:scatt_matrix}
\hat S(\nu)=\left( \begin{array}{cc} S_{11}(\nu)&S_{12}(\nu)\\
S_{21}(\nu)&S_{22}(\nu)\end{array} \right) \mbox{,}
\end{equation}
relating the amplitudes of the incoming and outgoing waves of
frequency $\nu$ in both infinite edges (leads). It should be emphasized that it is customary for microwave systems to make measurements of the scattering matrices in a function of microwave frequency $\nu$ which is related to the real part of the wave number  $\mathrm{Re\,}k=\frac{2\pi }{c}\nu$.

To measure the two-port scattering matrix $\hat S(\nu)$  the
vector network analyzer (VNA) Agilent E8364B was connected to
the microwave networks shown in Fig.~1(b) and Fig.~1(d).
The microwave test cables connecting
 microwave networks to the VNA are equivalent to attaching of
two infinite leads $e_1$ and $e_2$  to quantum graphs in Fig.~1(a) and Fig.~1(c).

\subsection{The two-edge network}
 The internal edge lengths of the two-edge network (see Figs.~1(a-b)) were parameterized by the parameter $t$ as $\ell_3 = \ell_0(1-t)$ and $\ell_4 = \ell_0$, with $\ell_0=1.0068\pm 0.0002$ m. The length of the edge $e_3$ was changed using microwave cables and a microwave phase shifter. The eigenvalue for $t=0$ is given by  $k=\frac{2\pi}{\ell_0}$ which in the frequency domain defines the resonance at 0.2978 GHz. Therefore, in order to analyze the dynamics of the topological resonance  in a function of the parameter $t$ the scattering matrix  $\hat S(\nu)$ of the network was measured in the frequency range $\nu = 0.01-0.5$ GHz.

 As an example, the modulus of the
determinant of the scattering matrix $|\det\bigr(\hat S(\nu)\bigl)|$ of the two-edge network for $t=-0.2$  is shown in Fig.~2(a) in the frequency range $0.30 - 0.36$ GHz (open circles). For $t\neq 0$ we deal for this network with two nearly-degenerated resonances $r_m=\nu_m+ig_m$, $m=1,2$. Therefore,  the parameters of the resonances, including  real $\mathrm{Re}\,k = \frac{2\pi}{c}\nu_1$ and imaginary $\mathrm{Im}\,k= \frac{2\pi}{c}g_1$ parts of the topological resonance, were obtained from the fit of the modulus of a sum of two Lorentzian functions \cite{Moldover1999}
\begin{equation}
\label{Eq:2_Lorentz}
f_2(\nu)= \sum^{2}_{m=1}\frac{i\nu A_{m}}{\nu^{2}-(\nu_{m}+ig_{m})^{2}}+B(\nu-\nu_{1})+C
\end{equation}
to the modulus of the determinant of the scattering matrix $|\det\bigr(\hat S(\nu)\bigl)|$,
where $A_m$, $B$, and $C$ are complex constants and  $r_m=\nu_m+ig_m$, $m=1,2$, are frequencies of the complex nearly-degenerated resonances.
 The fit of $|f_2(\nu)|$ (see Eq. \ref{Eq:2_Lorentz}) to the modulus   $|\det\bigr(\hat S(\nu)\bigl)|$ in the frequency range $\nu = 0.314-0.347$ GHz is marked in Fig.~2(a) by the red line. The topological resonance of the network is marked with a red dot and the other resonance with a blue dot. The right vertical axis $g$ in Fig.~2(a) shows the imaginary part of the resonances in GHz.

In Fig.~3 full circles show the trajectory of the topological resonance obtained experimentally for the two-edge network.
Even for the parameter $t=0$ the imaginary part of the experimental topological resonance  $g_1 = -43 \pm 20$ kHz is different than 0  suggesting that the topological resonance is influenced by the intrinsic absorption of the network. To analyze this situation we performed the numerical calculations using the method of pseudo-orbits \cite{BHJ,Li6,Li7}.
In the calculations we took into account the internal absorption of the microwave cables forming the edges of the microwave network. To do this we replaced the real wave vector $k$  by the complex one  with absorption-dependent imaginary part $\mathrm{Im\,}k=\beta \sqrt{2\pi \nu/c }$ and the real part $\mathrm{Re\,}k =  2\pi \nu/c $, where $\beta=0.009\,\mathrm{m}^{-1/2}$ is the absorption coefficient and $c$ is the speed of light in vacuum. This method is described in details in Ref. \cite{Hul2004}.

The results of the calculations are shown with diamonds in Fig.~3. The agreement between  the experimental results (full circles) and the numerical ones (diamonds) is very good showing that the non-zero value of the imaginary part of the  topological resonance at $t=0$ is due to intrinsic absorption in the network.

Due to the presence of the intrinsic absorption we fitted the experimental dependence of $\mathrm{Im}\,k$ on $t$ to the function $\mathrm{Im\,}k = a t^2 +b$ (see inset in Fig.~3). Using 9 experimental points (the central point corresponding to the topological resonance and four points to the left and four to the right from it) we obtained the values $a_{\mathrm{exp}} =  -2.11 \pm 0.40 \,\mathrm{m}^{-1}$ and  $b =  -0.00097 \pm 0.00051 \,\mathrm{m}^{-1}$. In the inset in Fig.~3 the theoretical fit is marked by the full red line.
 The experimental value $a_{\mathrm{exp}} =  -2.11 \pm 0.40 \,\mathrm{m}^{-1}$ is within the experimental error in agreement with the theoretical one $a_{\mathrm{th}} = -\frac{\pi^2}{4\ell_0}  = -2.45\,\mathrm{m}^{-1}$ obtained for $\ell_0 = 1.0068 \pm 0.0002 \,\mathrm{m}$. Moreover, the value  $b =  -0.00097 \pm 0.00051  \,\mathrm{m}^{-1}$ ($-46 \pm 24$ kHz) is in agreement with the imaginary part of the experimental topological resonance  $g_1 = -43 \pm 20$ kHz.

\subsection{The five-edge network}

In the case of the five-edge network the internal edge lengths of the network (see Figs.~1(c-d)) were parameterized by the parameter $t$ as
 $\ell_3 = \ell_0(1-t)$, $\ell_4 = \ell_0(1+t)$, $\ell_5 = \ell_0(1-t)$, $\ell_6 = \ell_0(1+t)$, and  $\ell_7 = \ell_0(1+t)$, with $\ell_0=1.0025\pm 0.0002$ m. The lengths of the edges $e_3, e_4, e_5$, and $e_6$ were changed using microwave cables and microwave phase shifters. The eigenvalue for $t=0$ can be found from the equation $k\ell_0 = \arccos{(-1/3)} = 1.9154$ \cite{LeeZworski16}. In the frequency domain it specifies the resonance  at 0.0912 GHz. That is why to analyze the dynamics of the topological resonance  in a function of the parameter $t$ the scattering matrix  $\hat S(\nu)$ of the five-edge network was measured in the frequency range $\nu = 0.01-0.5$ GHz. Fig.~2(b) shows the modulus of the
determinant of the scattering matrix $|\det\bigr(\hat S(\nu)\bigl)|$ of the five-edge network for $t=-0.05$ in the frequency range $0.06 - 0.12$ GHz (open circles). Here, the situation is even more complicated than in the case of the two-edge network because we deal with a structure of three nearly-degenerated resonances, with the topological resonance placed between the other two. That is why the parameters of the resonances, including  real $\mathrm{Re}\,k = \frac{2\pi}{c}\nu_2$ and imaginary $\mathrm{Im}\,k= \frac{2\pi}{c}g_2$ parts of the topological resonance, were obtained from the fit of the modulus of a sum of three Lorentzian functions

\begin{equation}
\label{Eq:3_Lorentz}
f_3(\nu) = \sum^{3}_{m=1}\frac{i\nu A_{m}}{\nu^{2}-(\nu_{m}+ig_{m})^{2}}+B_1(\nu-\nu_{1})+ B_2(\nu-\nu_{2}) +C
\end{equation}
 to the modulus  of the determinant of the scattering matrix $|\det\bigr(\hat S(\nu)\bigl)|$,
where $A_m$, $B_{1,2}$, and $C$ are complex constants and  $r_m=\nu_m+ig_m$, $m=1, 2, 3$, are frequencies of the complex nearly-degenerated resonances.
The fit of $|f_3(\nu)|$  to the modulus  $|\det\bigr(\hat S(\nu)\bigl)|$ in the frequency range $\nu = 0.074-0.116$ GHz for $t=-0.05$ is denoted in Fig.~2(b) by the red line. The topological resonance of the network is marked by a red dot while the two other ones by blue dots. The right vertical axis $g$ in Fig.~2(b) shows the imaginary part of the resonances in GHz.

The trajectory of the topological resonance obtained experimentally for the five-edge network (full circles) is shown in Fig.~4. In this case the departure from 0 of the imaginary part of the topological resonance $g_2=-0.55\pm 0.04$ MHz at $t=0$ is even more significant than in the case of the two-edge microwave network. The agreement of our numerical calculations (diamonds in Fig.~4) with the experimental results (full circles) clearly demonstrates that also in this case we deal with the effect of the internal absorption in the network. The fitted  dependence $\mathrm{Im\,}k = a t^2+b$ to the experimental points is marked by the full red line in the inset in Fig~4.
 Using 5 experimental points (the central point, two points to the left and two point to the right) we obtained the value $a_{\mathrm{exp}} = -0.46 \pm 0.03\,\mathrm{m}^{-1}$ and  $b =-0.0113 \pm 0.0002\,\mathrm{m}^{-1}$.
Within the experimental error the value $a_{\mathrm{exp}} = -0.46 \pm 0.03\,\mathrm{m}^{-1}$  corresponds to the theoretical one obtained for $\ell_0 = 1.0025 \pm 0.0002\,\mathrm{m}$ (this is the average of edge lengths $\ell_3$, $\ell_4$, $\ell_5$, and $\ell_6$ for $t=0$):
\begin{equation}
  a_{\mathrm{th}} = \frac{1}{2}\mathrm{Im\,}\ddot k = -\frac{1}{2\cdot 1.0025}[(2^2+(-2)^2)0.1711+2(-2)0.1141] \,\mathrm{m}^{-1} = -0.46\,\mathrm{m}^{-1}\,.
\end{equation}
Also in this case the value  $b=-0.011 \pm 0.001\,\mathrm{m}^{-1}$ ($-0.54 \pm 0.05$ MHz) is in agreement with the imaginary part of the experimental topological resonance  $g_2=-0.55 \pm 0.04$ MHz.

\section{Summary}

Using microwave networks with preserved time reversal invariance we investigated experimentally a Fermi golden rule which  gives rates of decay of states obtained by perturbing embedded eigenvalues of graphs and networks. We show that for the two-edge and five-edge microwave networks the embedded eigenvalues are connected with the topological resonances of the systems. Microwave networks are characterized by the intrinsic absorption which was taken into account in the numerical simulations of the networks. We found the trajectories of the topological resonances in the complex plane and showed that the experimental values of the parameter $a$ in the formula  $\mathrm{Im}\,k=a t^2 + b$ are very close to the expected theoretical ones in the formula  $\mathrm{Im\,}k = a_{\mathrm{th}} t^2$ for the two-edge and five-edge graphs, respectively. We show that the constant $b$ in the formula  $\mathrm{Im}\,k=a t^2 + b$ accounts for the intrinsic absorption of the networks. Although, we illustrated a Fermi rule for two particular graphs, the Theorem (1) from Ref.~\cite {LeeZworski16}  holds true for all quantum graphs with standard coupling conditions and at least one infinite lead, provided they support an eigenvalue embedded into the continuous spectrum for some edge lengths. It should be possible to obtain similar results for the corresponding microwave networks.

\section{Acknowledgements}
This work was supported in part by the National Science Center, Poland, Grant No. UMO-2018/30/Q/ST2/00324. J. L. was supported by the research programme ``Mathematical Physics and Differential Geometry'' of the Faculty of Science of the University of Hradec Kr\'alov\'e and the grant No. 18-00496S of the Czech Science Foundation.

\section{Appendix}

\subsection{Theoretical results for a Fermi rule. A case of a two-edge graph}

Let us consider a graph consisting of two vertices, two internal and two external edges (see Fig.~1(a)). Let the lengths of the internal edges be $\ell_3<\infty$ and $\ell_4 <\infty$, while the edges $e_1$ and $e_2$ have infinite lengths. Let us consider the dependence of the edge lengths on the parameter $t$ as $\ell_3 = \ell_0(1-t)$, $\ell_4 = \ell_0$ and the eigenvalue for $t=0$ with $k=\frac{2\pi}{\ell_0}$. This situation corresponds to the case in \cite[Fig.~2 (b)]{LeeZworski16}. Let the edges be parametrized from $x=0$ at $v_1$ to $x=\ell_3$ and $x=\ell_4$ at $v_2$. We find from the above expressions that
$$
  \dot a_3 = -\frac{1}{\ell_0}\frac{\partial \ell_3}{\partial t} = 1\,,\quad \dot a_4 = -\frac{1}{\ell_0}\frac{\partial \ell_4}{\partial t} = 0\,.
$$
The normalized eigenfunction for $t=0$ and $k = \frac{2\pi}{\ell_0}$ has edge components $u_1(x) = 0$, $u_2(x) = 0$, $u_3(x) = \frac{1}{\sqrt{\ell_0}}\sin{(kx)}$, $u_4(x) = - \frac{1}{\sqrt{\ell_0}}\sin{(kx)}$.

Let us now compute the form of the generalized eigenfunctions $e^s(k,x)$, $s=1,2$. The form of the components of $e^1(k,x)$ follows from its definition.
\begin{eqnarray*}
  e^1_1(k,x) = \mathrm{e}^{-\mathrm{i}kx}+s_{11}\mathrm{e}^{\mathrm{i}kx}\,,&\quad & e^1_2(k,x) = s_{12} \mathrm{e}^{\mathrm{i}kx}\,,\\
  e^1_3(k,x) = \alpha_3\sin{(kx)}+\beta_3\cos{(kx)}\,,& \quad & e^1_4(k,x) = \alpha_4\sin{(kx)}+\beta_4\cos{(kx)}
\end{eqnarray*}
with unknown constants $s_{11}$, $s_{12}$, $\alpha_3$, $\beta_3$, $\alpha_4$, $\beta_4$.
The coupling conditions (\ref{eq:cc}) yield
\begin{eqnarray*}
  1+s_{11} = \beta_3 = \beta_4\,,\quad \mathrm{i}(-1+s_{11})+\alpha_3+ \alpha_4 = 0\,,\\
  s_{12} = \alpha_3 \sin{(k\ell_3)} +\beta_3\cos{(k\ell_3)} = \alpha_4 \sin{(k\ell_4)} +\beta_4\cos{(k\ell_4)}\,,\\
  \mathrm{i} s_{12}  -\alpha_3 \cos{(k\ell_3)} +\beta_3\sin{(k\ell_3)} -\alpha_4 \cos{(k\ell_4)} +\beta_4\sin{(k\ell_4)} = 0\,.
\end{eqnarray*}
If one writes this set of six equations for six variables into the matrix form, one finds that its solutions are not properly defined for $\ell_3 = \ell_4 = \ell_0$ (i.e., the case $t=0$) since the determinant of the corresponding matrix is 0. However, following the definition of $e^s$, one can use the holomorphic extensions of the solutions to $k = \frac{2\pi}{\ell_0}$ and obtain the unknown coefficients as the limits for $t \to 0$. We find
$$
  \alpha_3 = \alpha_4 = \frac{\mathrm{i}}{2}\,,\quad \beta_3 = \beta_4 = 1\,,\quad s_{11} = 0\,,\quad s_{12} = 1\,.
$$
This corresponds to
$$
  e^1_1(k,x) = \mathrm{e}^{-\mathrm{i}kx}\,,\quad e^1_2(k,x) = \mathrm{e}^{\mathrm{i}kx}\,,\quad e^1_3 (k,x) = e^1_4 (k,x) = \cos{(kx)}+\frac{\mathrm{i}}{2}\sin{(kx)}\,.
$$

We proceed similarly for $e^2(k,x)$. Using the ansatz
\begin{eqnarray*}
  e^2_1(k,x) = s_{21}\mathrm{e}^{\mathrm{i}kx}\,,&\quad & e^2_2(k,x) = \mathrm{e}^{-\mathrm{i}kx} + s_{22}\mathrm{e}^{\mathrm{i}kx}\,,\\
  e^2_3(k,x) = \gamma_3 \sin{(kx)}+ \delta_3 \cos{(kx)}\,,& \quad &  e^2_4(k,x) = \gamma_4 \sin{(kx)}+ \delta_4 \cos{(kx)}
\end{eqnarray*}
the coupling conditions (\ref{eq:cc}) yield the set of equations
\begin{eqnarray*}
  s_{21} = \delta_3 = \delta_4\,,\quad \mathrm{i}s_{21}+\gamma_3+\gamma_4 = 0\,,\\
  1+s_{22} = \gamma_3 \sin{(k\ell_3)} + \delta_3 \cos{(k\ell_4)} =  \gamma_4 \sin{(k\ell_4)} + \delta_4 \cos{(k\ell_4)}\,,\\
  \mathrm{i}(-1+s_{22}) -\gamma_3\cos{(k\ell_3)} + \delta_3 \sin{(k\ell_3)} -\gamma_4\cos{(k\ell_4)} + \delta_4 \sin{(k\ell_4)} = 0\,.
\end{eqnarray*}
The solutions after the holomorphic extension to $k = \frac{2\pi}{\ell_0}$ are
$$
  s_{21} = \delta_3 = \delta_4 = 1\,,\quad \gamma_3 = \gamma_4 = -\frac{\mathrm{i}}{2}\,,\quad s_{22} = 0\,.
$$
This corresponds to
$$
  e^2_1(k,x) = \mathrm{e}^{\mathrm{i}kx}\,,\quad e^2_2(k,x) = \mathrm{e}^{-\mathrm{i}kx}\,,\quad e^2_3 (k,x) = e^2_4 (k,x) = \cos{(kx)}-\frac{\mathrm{i}}{2}\sin{(kx)}\,.
$$
Hence
\begin{eqnarray*}
  \left<\dot a u(x),e^1(k,x)\right> = \frac{1}{\sqrt{\ell_0}}\int_0^{\ell_0} \sin{(kx)}\left(\cos{(kx)}+\frac{\mathrm{i}}{2}\sin{(kx)}\right)\,\mathrm{d}x = \frac{\mathrm{i}}{4}\sqrt{\ell_0}\,,\\
  \left<\dot a u(x),e^2(k,x)\right> = \frac{1}{\sqrt{\ell_0}}\int_0^{\ell_0} \sin{(kx)}\left(\cos{(kx)}-\frac{\mathrm{i}}{2}\sin{(kx)}\right)\,\mathrm{d}x = - \frac{\mathrm{i}}{4}\sqrt{\ell_0}\,.
\end{eqnarray*}
Moreover, using
\begin{eqnarray*}
  \partial_\nu u_3(v_1) = -\frac{k}{\sqrt{\ell_0}} = -\frac{2\pi}{\ell_0^{3/2}}\,,\quad \partial_\nu u_4(v_1) = \frac{k}{\sqrt{\ell_0}} = \frac{2\pi}{\ell_0^{3/2}}\,,\\
  \partial_\nu u_3(v_2) = \frac{k}{\sqrt{\ell_0}}\cos{(k\ell_0)} = \frac{2\pi}{\ell_0^{3/2}}\,,\quad \partial_\nu u_4(v_2) = -\frac{k}{\sqrt{\ell_0}}\cos{(k\ell_0)} = -\frac{2\pi}{\ell_0^{3/2}} \,,\\
  e^1(k,v_1) = e^1(k,v_2) =e^2(k,v_1) = e^2(k,v_2) = 1 \,,\quad u(v_1) = u(v_2) = 0
\end{eqnarray*}
we can find
$$
  \sum_v\sum_{e_j\ni v}\frac{1}{4}\dot a_j[3\partial_\nu u_j(v) \overline{e^s(k, v)}-u(v)\partial_\nu\overline{e^s_j(k,v)}] = \frac{1}{4}\cdot 1\cdot \left[3\cdot \left(-\frac{2\pi}{\ell_0^{3/2}}\right) - 0\right]+\frac{1}{4}\cdot 1\cdot \left[3\cdot \frac{2\pi}{\ell_0^{3/2}} - 0\right] = 0\,.
$$
Hence, only the first terms of the functions $F_s$ are nonzero.
$$
  |F_1| = \left|\frac{2\pi}{\ell_0}\frac{i}{4}\sqrt{\ell_0}\right| = \frac{\pi}{2\sqrt{\ell_0}} \,,\quad |F_2| = \left|\frac{2\pi}{\ell_0}\frac{(-i)}{4}\sqrt{\ell_0}\right| = \frac{\pi}{2\sqrt{\ell_0}}
$$
and
$$
  \mathrm{Im\,}\ddot k = -\frac{\pi^2}{2\ell_0}\,.
$$
One can simply prove (see, e.g. \cite{LeeZworski16,ExnerLipovsky17}) that $\mathrm{Im}\,\dot k|_{t=0} = 0$ and clearly $\mathrm{Im}\,k|_{t=0} = 0$. Therefore, we see from the Taylor's expansion that the imaginary part of $k(t)$ near the eigenvalue behaves as
$$
  \mathrm{Im}\,k \approx -\frac{\pi^2}{4\ell_0} t^2\,.
$$

\pagebreak

\begin{figure}[tb]
\includegraphics[width=0.8\linewidth]{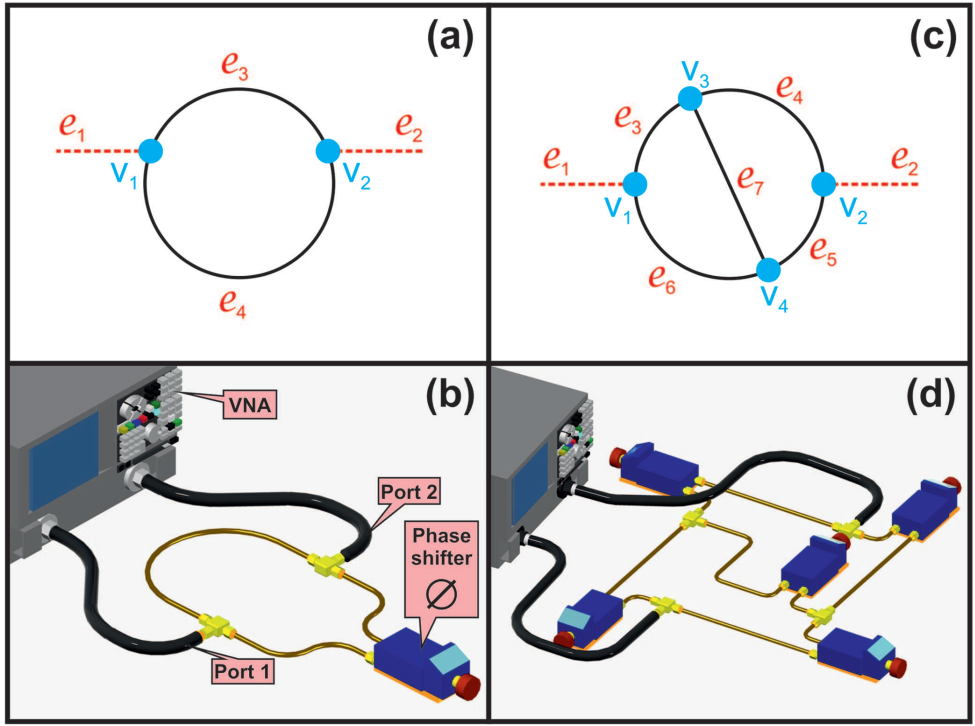}
\caption{
Panels (a) and (b) show the schemes of  a two-edge quantum graph with $\mathcal{V}=2$ vertices and a  microwave network  with the same topology. Panels (c) and (d) show the schemes of  a five-edge quantum graph with $\mathcal{V}=4$ vertices and a  microwave network  with the same topology.
 The microwave networks were connected to the vector network analyzer with the flexible microwave cables which are equivalent to attaching  infinite leads  to  quantum graphs in panels (a) and (c).
}
\label{Fig1}
\end{figure}

\begin{figure}[tb]
\includegraphics[width=0.6\linewidth]{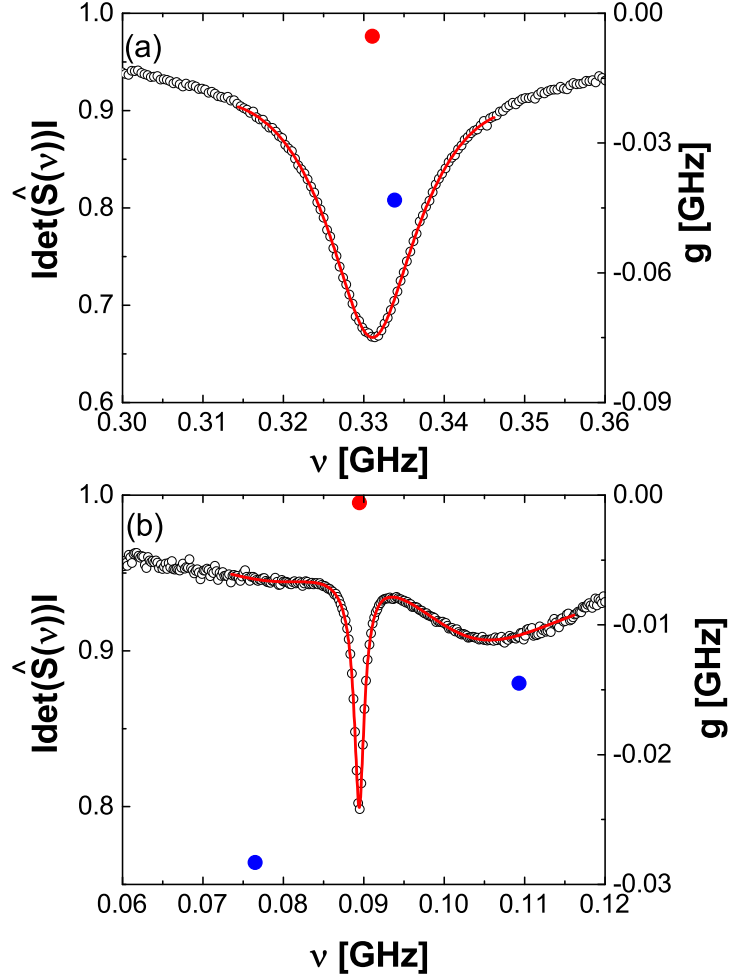}
\caption{(a) The modulus $|\det\bigr(\hat S(\nu)\bigl)|$ of the
determinant of the scattering matrix of the two-edge network for the parameter $t=-0.2$ in the frequency range $0.30 - 0.36$ GHz (open circles).
The fit of $|f_2(\nu)|$ (see Eq. \ref{Eq:2_Lorentz}) to the modulus   $|\det\bigr(\hat S(\nu)\bigl)|$  in the frequency range $\nu = 0.314-0.347$ GHz is marked  by the red line. The topological resonance of the network is marked with a red dot and the other one with a blue dot.  The right vertical axis $g$ shows the imaginary part of the resonances in GHz.
(b) The modulus $|\det\bigr(\hat S(\nu)\bigl)|$ of the
determinant of the scattering matrix of the five-edge network for the parameter $t=-0.05$ in the frequency range $0.06 - 0.12$ GHz (open circles).
The fit of $|f_3(\nu)|$ (see Eq. \ref{Eq:3_Lorentz}) to the modulus  $|\det\bigr(\hat S(\nu)\bigl)|$ in the frequency range $\nu = 0.074-0.116$ GHz is denoted by the red line. The topological resonance of the network is marked by a red dot while the two other ones by blue dots. The right vertical axis $g$ shows the imaginary part of the resonances in GHz.
}
\label{Fig2}
\end{figure}

\begin{figure}[tb]
\includegraphics[width=0.8\linewidth]{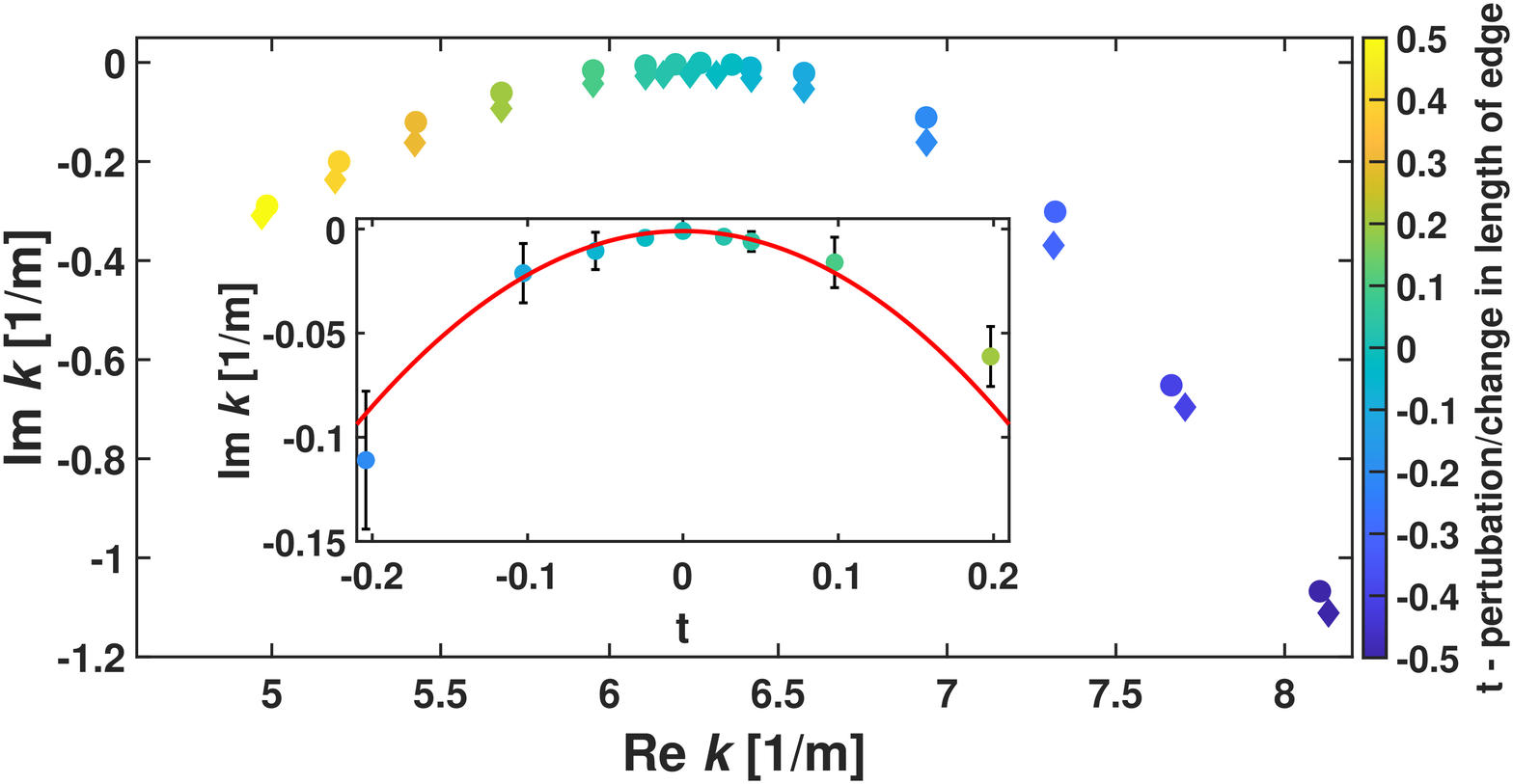}
\caption{
The trajectory of the topological resonance obtained experimentally for the two-edge network (full circles). The numerical calculations taking into account the intrinsic absorption of the network are marked by diamonds. The color coding on the right vertical axis indicates the parameter $t$.
In the inset we show the fitted dependence  $\mathrm{Im\,}k = a t^2+b$  (red line) to the experimental points (full circles). Using 9 experimental points (the central point corresponding to the topological resonance and four points to the left and four to the right from it)  the values $a_{\mathrm{exp}} =  -2.11 \pm 0.40 \,\mathrm{m}^{-1}$ and $b =-0.00097 \pm 0.00051 \,\mathrm{m}^{-1}$ were obtained.  The experimental value $a_{\mathrm{exp}}$ corresponds within the experimental error to the theoretical value $a_{\mathrm{th}} = -\frac{\pi^2}{4\ell_0}  = -2.45\,\mathrm{m}^{-1}$.
}
\label{Fig3}
\end{figure}

\begin{figure}[tb]
\includegraphics[width=0.8\linewidth]{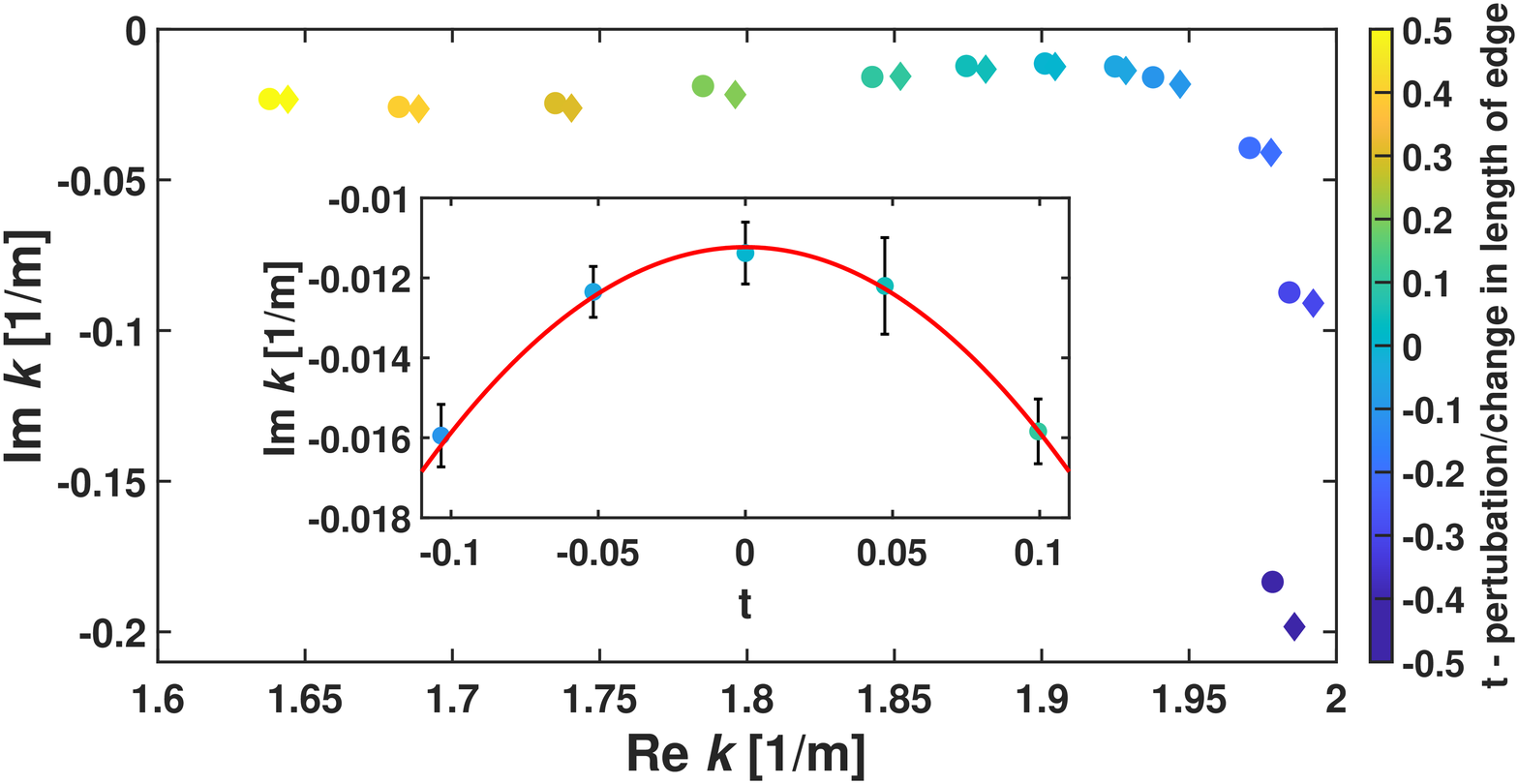}
\caption{
The trajectory of the topological resonance obtained experimentally for the five-edge network (full circles).
The numerical calculations taking into account the intrinsic absorption of the network are marked by diamonds. The color coding on the right vertical axis indicates the parameter $t$.
In the inset we show the fitted  dependence  $\mathrm{Im\,}k = a t^2+b$  (red line) to the experimental points (full circles).
Using 5 experimental points (the central point with $\mathrm{Re\,}k \simeq 1.9 \mathrm{\, m}^{-1}$, two points to the left and two point to the right) we obtained the values $a_{\mathrm{exp}} = -0.46 \pm 0.03\,\mathrm{m}^{-1}$ and  $b = -0.0113 \pm 0.0002\,\mathrm{m}^{-1}$. The experimental value $a_{\mathrm{exp}}$ corresponds within the experimental error to the theoretical value  $a_{\mathrm{th}} = -0.46\,\mathrm{m}^{-1}$.
}
\label{Fig4}
\end{figure}

\end{document}